\documentclass[prl,aps,showpacs,twocolumn,unsortedaddress]{revtex4}

\usepackage{graphics, bm}
\usepackage{psfrag}
\usepackage{amsmath}
\usepackage{amssymb}
\usepackage{epsfig}

\newcommand{\beq}    {\begin{equation}}
\newcommand{\enq}    {\end{equation}}

\newcommand{\kk}     {{\rm \bf k }}
\newcommand{\rr}     {{\rm \bf r }}
\newcommand{\rR}     {{\rm \bf R }}

\newcommand{\df}     {\equiv}

\begin{document}

\title{Spatially dependent Kondo effect in Quantum Corrals}

\author{Enrico Rossi and Dirk K. Morr}
\affiliation{Department of Physics, University of Illinois at Chicago, Chicago IL 60607}

\date{\today}


\begin{abstract}

We study the Kondo screening of a single magnetic impurity inside a
non-magnetic quantum corral located on the surface of a metallic
host system. We show that the spatial structure of the corral's
eigenmodes lead to a spatially dependent Kondo effect whose
signatures are spatial variations of the Kondo temperature, $T_K$.
Moreover, we predict that the Kondo screening is accompanied by the
formation of multiple Kondo resonances with characteristic spatial
patterns. Our results open new possibilities to manipulate and
explore the Kondo effect by using quantum corrals.

\end{abstract}




\maketitle


The Kondo effect exhibited by a magnetic impurity is one of the most
fundamental and important phenomena in condensed matter physics
\cite{Kon64,Wil75,Hew97}. Over the last few years, the emergence of
a Kondo effect in confined host geometries with discrete energy
levels, such as quantum dots \cite{dots}, nanotubes \cite{tubes} and
molecules \cite{Boo05} has attracted significant experimental
\cite{dots,tubes,Boo05} and theoretical \cite{Thi99,Sch02} interest.
Discrete eigenmodes have also been observed in {\it quantum corrals}
located on metallic surfaces \cite{Cro95}. Recently, Manoharan {\it
et al.}~using scanning tunneling microscopy (STM) showed that the
spatial structure of these eigenmodes can be employed to create the
quantum image of a Kondo resonance \cite{Man00}. While a series of
theoretical studies successfully explained the formation of this
quantum image \cite{theory}, the question whether the eigenmodes'
spatial structure leads to a spatially dependent Kondo effect for
different positions of a magnetic impurity inside the quantum
corral, has not yet been addressed.

In this Letter, we answer this question by studying the Kondo
screening of a single magnetic impurity inside a non-magnetic
quantum corral located on the surface of a metallic host. Combining
a {\it large-$N$} expansion \cite{Hew97,Read83,Bic87} with a
generalized ${\hat T}$-matrix approach \cite{Morr04} we show that
the spatial structure of the corral's eigenmodes leads to a
spatially dependent Kondo effect whose signatures are spatial
variations of the Kondo temperature, $T_K$, and of the critical
coupling, $J_{cr}$. Specifically, $T_K$, is the largest and $J_{cr}$
the smallest, at those locations where the density of states (DOS)
of the lowest energy eigenmode possesses a maximum. Moreover, we
find that the screening of the magnetic impurity leads to the
formation of multiple Kondo resonances with characteristic spatial
patterns that provide clear experimental signatures of the spatially
dependent Kondo effect. Our results demonstrate that quantum corrals
provide a new possibility to manipulate and explore the nature of
the Kondo effect.

While the {\em large-N} approach \cite{Hew97,Read83,Bic87} provides
a qualitatively correct description of all salient features of the
Kondo effect \cite{Wil75}, its interpretation requires some care. In
particular, in the {\it large-N} approach, the onset of Kondo
screening occurs via a sharp transition, such that for a given $J$
($T$), a Kondo effect occurs for $T<T_K$ ($J>J_{cr}$). This sharp
transition is an artifact of the {\it large-N} approach
\cite{Wil75}, and hence $T_K$ and $J_{cr}$ should rather be
interpreted as crossover values. However, this interpretation does
not change the main result of our study, {\it viz.} $T_K$ and
$J_{cr}$ exhibit a pronounced spatial dependence with clear
experimental signatures.

We consider a system consisting of a quantum corral with $M$
non-magnetic impurities located on the surface of a metallic host,
and a single magnetic impurity inside the corral. This system is
described by the Hamiltonian
\begin{align}
 H = &-\sum_{{\bf i,j},\sigma} t_{\bf ij} \ c^\dagger_{{\bf i},\sigma} c_{{\bf j},\sigma} +
      U {\sum_{i,\sigma}}^\prime c^\dagger_{{\bf r}_i,\sigma}c_{{\bf r}_i,\sigma} \nonumber \\
     & \quad + J{\bf S}\cdot
     c^\dagger_{ \rR \sigma}{\bf\tau}_{\sigma\sigma'}c_{\rR \sigma'}
     \ ,
 \label{eq:ham}
\end{align}
where $c^\dagger_{{\bf i},\sigma} c_{{\bf i},\sigma}$ are the
fermionic creation and annihilation operators for a (host)
conduction electron at site ${\bf i}$ with spin $\sigma$, and
$t_{\bf ij}$ is the hopping element between sites ${\bf i}$ and
${\bf j}$. In the following, we consider a two-dimensional (2D) host
metal on a square lattice with dispersion $\epsilon_\kk=k^2/2m-\mu$
where $\mu$ is the chemical potential. We set the lattice constant
$a_0$ to unity and use $E_0\df\hbar^2/ma_0^2$ as our unit of energy.
The primed sum runs over all positions ${\bf r}_i \ (i=1,..,M)$ of
the $M$ corral impurities with identical non-magnetic scattering
potential $U$. The last term in Eq.(\ref{eq:ham}) describes the
Kondo interaction between the conduction electrons and the magnetic
impurity, located at site $\rR$ with magnetic moment ${\bf S}$ and
scattering strength $J$. Below, we consider for concreteness a
magnetic impurity with spin $S=1/2$.

In the {\em large-N} approach \cite{Hew97,Read83,Bic87}, the spin ${
\bf S}$ of the magnetic impurity is expressed in terms of fermionic
operators, $f^\dagger_m, f_m$, that obey the constraint
$\sum_{m=1..N} f^\dagger_m f_m=1$ with $N=2$ being the number of
fermionic flavors for a magnetic impurity with spin $S=1/2$. Within
a path integral approach, the constraint is enforced by means of a
Lagrange multiplier $\epsilon_f$ and the exchange interaction in
Eq.(\ref{eq:ham}) is decoupled via a Hubbard-Stratonovich field,
$s$. $\epsilon_f$ is interpreted as the energy of the $f$ electrons,
and $s^2$ represents their hybridization with the conduction
electrons. By minimizing the effective action on the saddle point
level, one obtains the self-consistent equations \cite{com1}
\begin{subequations}
\begin{eqnarray}
T \sum_n \frac{1}{i \omega_n -\epsilon_f - s^2 G_c(\rR, \rR , i\omega_n)}+\frac{1}{2}&=&\frac{1}{N} \label{eq:saddle1} \\
T \sum_n \frac{G_c(\rR, \rR ,i\omega_n)}{i \omega_n -\epsilon_f -
s^2 G_c(\rR , \rR,i\omega_n)}&=&\frac{1}{J}. \label{eq:saddle2}
\end{eqnarray}
\end{subequations}
$G_c$, the conduction electrons' Green's function in the presence of
the corral only, is given by \cite{Morr04}
\begin{align}
 G_c(\rr,\rr', i\omega_n)  = & G_0(\rr - \rr', i\omega_n) +
 {\sum_{j,l}}^\prime G_0(\rr - \rr_j, i\omega_n)    \nonumber \\
                        &  \hspace{-1cm} \times \ T_{jl}(i\omega_n) G_0(\rr_l - \rr', i\omega_n)
\label{eq:G_c}
\end{align}
where $G_0=1/(i\omega_n-\epsilon_{\bf k})$ is the Green's function
of the unperturbed host system in momentum space. The ${\hat
T}$-matrix is obtained from the Bethe-Salpeter equation
\beq
 T_{ij}(i \omega_n) = U \delta_{ij}  +
                         U {\sum_{l}}^\prime G_0(\rr_i - \rr_l,i \omega_n) T_{li}(i \omega_n).
 \label{eq:b-s}
\enq
In the presence of the magnetic impurity, the total Green's function
of the conduction electrons is given by
\begin{align}
 G_c^{tot}(\rr,\rr',i \omega_n) = &G_c(\rr,\rr',i\omega_n) + \nonumber \\
                      & \hspace{-1cm}
                      s^2G_c(\rr,\rR,i\omega_n)F(i\omega_n)G_c(\rR,\rr',i\omega_n)
                      \ ,
 \label{eq:G_tot}
\end{align}
where  $F=[i\omega_n - \epsilon_f -s^2 G_c({\bf R},{\bf R},i
\omega_n)]^{-1}$ is the Green's function of the $f$ electrons. In
the presence (absence) of the magnetic impurity, the host system's
DOS, $N_c^{tot}$ ($N_c$), is obtained from Eq.(\ref{eq:G_tot})
[Eq.(\ref{eq:G_c})] via $N^{(tot)}_c({\bf r},\omega)=-2 {\rm Im}[
G_c^{(tot)}(\rr,\rr,\omega+i\delta)]/ \pi$ with $\delta=0.0025 E_0$.

\begin{figure}[!h]
 \begin{center}
  \includegraphics[width=8.5cm]{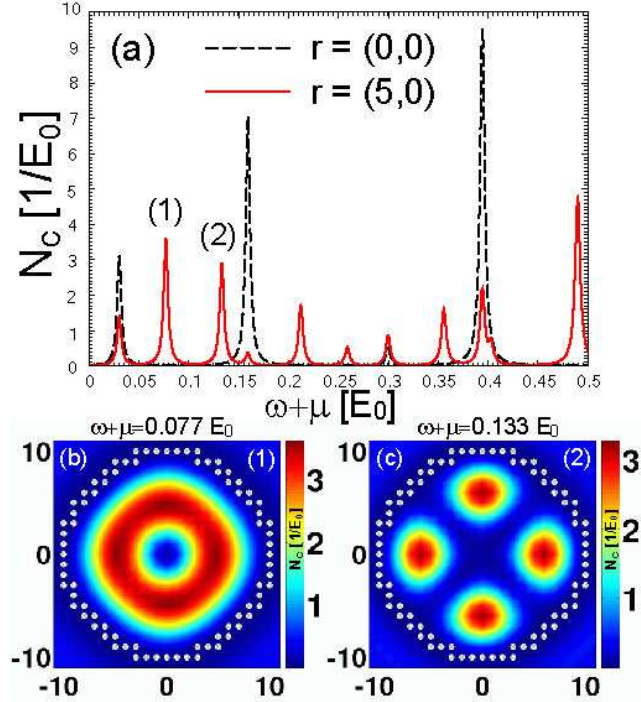}
  \caption{(Color online)
           (a) DOS, $N_c$, as a function of $\omega+\mu$ at $\rr_1 = (0,0)$ and $\rr_2 = (5,0)$.
       (b),(c) Spatial plot of $N_c$ at $\omega+\mu= 0.077 E_0$ and $0.133 E_0$,
       corresponding to modes (1) and (2) in (a), respectively. The filled white circles represent
       the non-magnetic impurities forming
               the quantum corral.
          }
  \label{fig:1}
 \end{center}
\end{figure}

In what follows, we consider a circular quantum corral of radius
$r=10 a_0$ consisting of $M=81$ non-magnetic impurities with $U=2.5
E_0$ [see Fig.~\ref{fig:1}(b)]. In the absence of the magnetic
impurity, the DOS, $N_c$, exhibits well separated eigenmodes
[Fig.~\ref{fig:1}(a)] that possess distinct spatial patterns, as
shown in Figs.~\ref{fig:1}(b) and (c) for the eigenmodes denoted by
$(1)$ and $(2)$ in Fig.~\ref{fig:1}(a).
%
%
We expect that the spatial structure of the Kondo effect is
determined by that of the lowest energy eigenmode as long as the
Kondo temperature, $T_K$, is smaller than the energy splitting
between modes. To test this conjecture, we set $\mu=0.077 E_0$, such
that the mode shown in Fig.~\ref{fig:1}(b) is located at the Fermi
energy.

\begin{figure}[!h]
 \begin{center}
  \includegraphics[width=8.5cm]{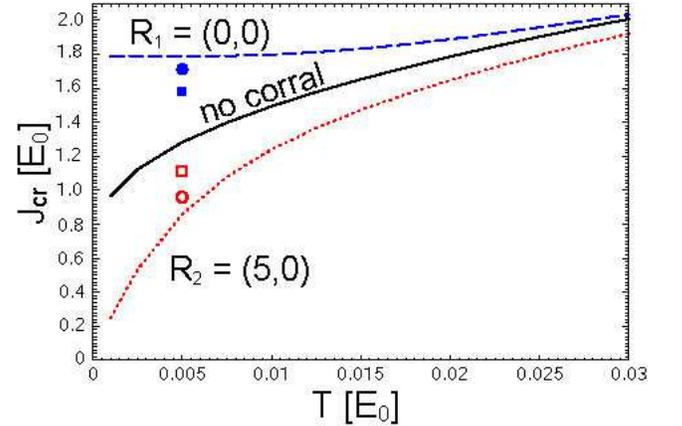}
  \caption{(Color online). $J_{cr}$ at
 $\rR_{1,2}$ as a function of $T$. The solid line represents $J^u_{cr}$ for a system without
 corral. The filled (open) square and circle represent $J_{cr}$ at $\rR_1$ ($\rR_2$) for
 a quantum corral with $U=0.2 E_0$ and $0.5 E_0$, respectively.}
 \label{fig:2}
 \end{center}
\end{figure}

In Fig.~\ref{fig:2} we present $J_{cr}$ \cite{com2} as a function of
temperature for two positions $\rR_{1,2}$ of the magnetic impurity,
corresponding to the minimum [$\rR_1=(0,0)$] and maximum
[$\rR_2=(5,0)$] in the DOS of the zero-energy mode.
%
%
In the limit $T \to 0$, the behavior of $J_{cr}$ at $\rR_{1,2}$ is
qualitatively different: while $J_{cr} \sim T $ at $\rR_2$, $J_{cr}$
saturates to a finite value at $\rR_1$. This results can be
understood by considering a  model in which the fermionic excitation
spectrum inside the corral consists of discrete eigenmodes,
described by the Green's function $G_c(\rr,i\omega_n)=\sum_l
\varphi_l(\rr)/(i\omega_n-\Omega_l)$, where $\Omega_l$ and
$\varphi_l(\rr)$ are the energy and the spectral weight of the
$l'th$ mode at position $\rr$, respectively. With this form of
$G_c$, Eq.(\ref{eq:saddle2}) yields \cite{com2}
\beq
 \frac{1}{J_{cr}} = -\sum_l\frac{\varphi_l(\rR)}{\Omega_l}
                     \left[n_F\left(\frac{\Omega_l}{T}\right) - \frac{1}{2}\right].
 \label{eq:J_cr_2}
\enq
Taking one mode to be located at zero energy, $\Omega_p=0$, while
for all other modes $|\Omega_l| \gg T$, one finds in the
low-temperature limit $\varphi_p(\rR)/4T\gg \sum_{l\neq
p}\varphi_l(\rR)/2|\Omega_l|$ that $J_{cr}\approx
4T/\varphi_p(\rR)$. This low-temperature behavior is observed at
$\rR_2$ since $\varphi_p(\rR_2) \not = 0$ for the zero-energy mode
shown in Fig.~\ref{fig:1}(b). In contrast, one has $\varphi_p(\rR_1)
= 0$, implying that for $T \rightarrow 0$, $J_{cr}(\rR_1)$ saturates
to a non-zero value given by $J_{cr} \approx 1/\sum_{l\neq
p}\varphi_l(\rR_1)/2|\Omega_l|$. In Fig.~\ref{fig:2}, we also plot
the temperature dependence of $J_{cr}$ for a system without quantum
corral, $J^u_{cr}$, which is qualitatively similar to that of
$J_{cr}(\rR_2)$ due to the non-zero DOS of the unperturbed, metallic
host at $\omega=0$. Since $J^u_{cr}$ lies between $J_{cr}(\rR_1)$
and $J_{cr}(\rR_2)$, a quantum corral can either facilitate (at
$\rR_2$) or suppress (at $\rR_1$) the screening of a magnetic
impurity. Finally, the difference in $J_{cr}$ between $\rR_1$ and
$\rR_2$ is quite substantial already for rather weak scattering
potential $U$, as follows from Fig.~\ref{fig:2} where the filled
[open] square and circle represent $J_{cr}$ at $\rR_1$ [$\rR_2$] for
a corral with $U=0.2 E_0$ and $0.5 E_0$, respectively. This
demonstrates the robustness of the spatially dependent Kondo effect
even for small $U$.

\begin{figure}[!h]
 \begin{center}
  \includegraphics[width=8.5cm]{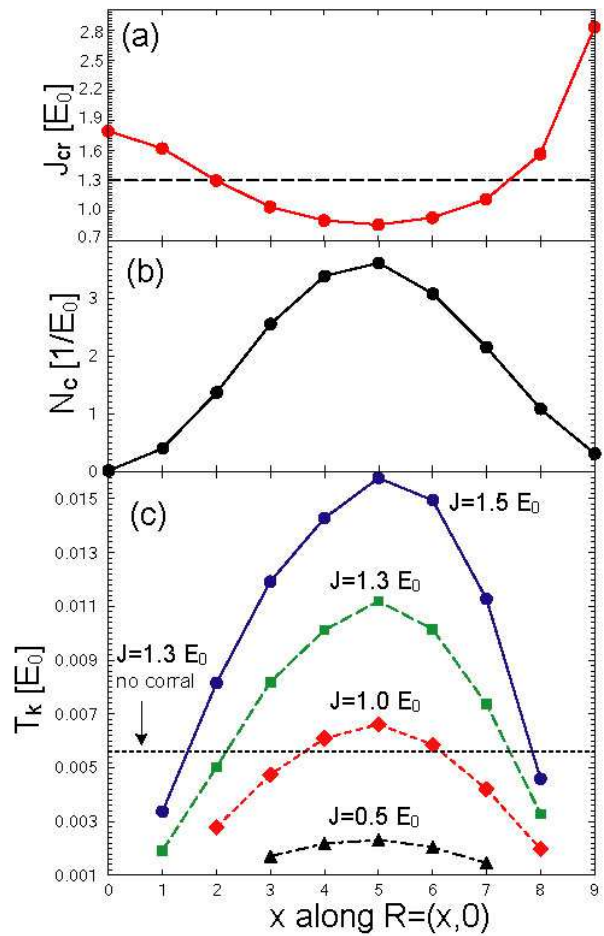}
  \caption{(Color online).
  (a) $J_{cr}$ along $\rR=(x,0)$ for $T= 0.005 E_0$. (b) $N_c$ at $\omega=0$ along the same path as in (a).
(c) $T_K$ for several $J$ along the same path as in (a). The dotted
line shows $T_K$ for a magnetic impurity with $J=1.3 E_0$ in the
absence of a corral. If no value for $T_K$ is given, $T_K$ is
smaller than the lowest temperature we considered, $T=10^{-3}E_0$.}
  \label{fig:3}
 \end{center}
\end{figure}

In Fig.~\ref{fig:3}(a) we present $J_{cr}$ along $\rR=(x,0)$ inside
the corral for $T=0.005 E_0$.
%
%
A comparison with the DOS of the zero-energy mode in
Fig.~\ref{fig:3}(b) along the same path shows that $J_{cr}(\rR)$
exhibits as expected a minimum at $\rR=(5,0)$ where the DOS
possesses a maximum. Complementary to this result, we plot in
Fig.~\ref{fig:3}(c) the spatial dependence of $T_K$, which exhibits
the same spatial dependence as (a) the DOS in Fig.~\ref{fig:3}(b),
and (b) the hybridization, $s^2$ for $T \rightarrow 0$ (not shown).
Hence, a magnetic impurity with a given $J$ exhibits characteristic
signatures of Kondo screening (see below), only for those locations
inside the corral, for which $T<T_K(\rR)$.
This result remains qualitatively unaffected by the interpretation
of $T_K$ as a crossover. A comparison of $T_K(\rR)$ with the
spatially uniform $T_K$ in the absence of a corral [dotted line in
Fig.~\ref{fig:3}(c) for $J=1.3 E_0$] shows that $T_K(\rR)$ is
increased inside the corral for $3 \leq x \leq 7$, and suppressed
otherwise. This opens the possibility to custom-design $T_K$ for a
magnetic impurity inside a quantum corral.

\begin{figure}[!h]
 \begin{center}
  \includegraphics[width=8.5cm]{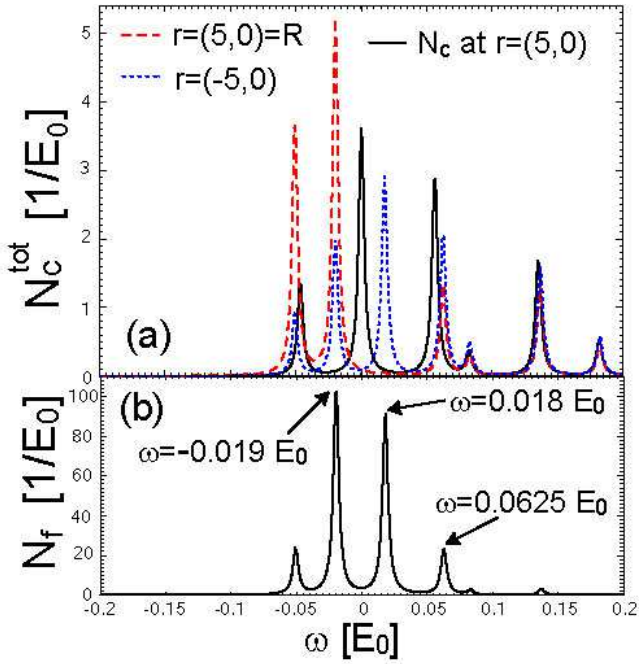}
  \caption{(Color online) (a) $N^{tot}_c$ and (b) $N_f$ as a function of frequency for
  $\rR=(5,0)$, $J=1.45 E_0$, $T=0.005 E_0$, $\epsilon_f=0.02\;E_0$ and $s^2=0.03\;E_0^2$.
          }
  \label{fig:4}
 \end{center}
\end{figure}

In Figs.~\ref{fig:4}(a) and (b), we present $N^{tot}_c$ and the $f$
electrons' DOS, $N_f=-N{\rm Im}[F(\rR,\omega)]/\pi$, respectively,
as a function of energy at $T=0.005 E_0$ for an impurity with
$J=1.45\;E_0$ and $\rR=(5,0)$, yielding $T_K=0.0145\;E_0$.
%
%
We find that Kondo screening is accompanied by multiple Kondo
resonances, a phenomenon characteristic of the discrete excitation
spectrum in a host system \cite{Thi99,Ali05}. A comparison of
Figs.~\ref{fig:4}(a) and (b) shows that away from the Fermi energy,
each corral eigenmode induces a single Kondo resonance, leading to a
small shift of the eigenmode energy from its unperturbed value. In
contrast, level repulsion between the unperturbed $f$-level and the
corral's zero-energy eigenmode leads to two Kondo resonances almost
symmetrically located around $\omega=0$. This splitting of the
corral's eigenmode into two Kondo resonances is one of the most
prominent signatures of Kondo screening in the DOS. Note that due to
the frequency dependence of $N_c$, the width of the low-energy Kondo
resonances is {\it not} set by $T_K$ \cite{Hew97}.

\begin{figure}[!h]
 \begin{center}
  \includegraphics[width=8.5cm]{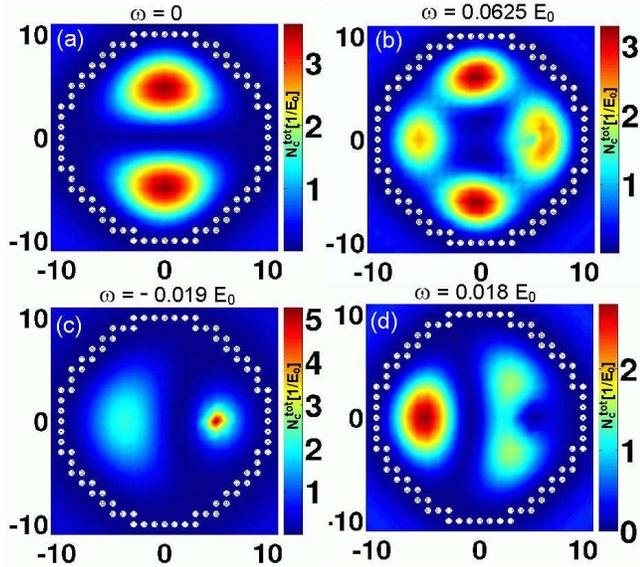}
  \caption{(Color online) Spatial form of $N_c^{tot}$ at (a)
  $\omega=0$ and at three of the Kondo resonances
  shown in Fig.~\ref{fig:4}: (b) $\omega =  0.0625\;E_0$,
           (c) $\omega = -0.019\;E_0$,
       (d) $\omega =  0.018\;E_0$.
          }
  \label{fig:5}
 \end{center}
\end{figure}

The spatial form of $N_c^{tot}$ at several frequencies is shown in
Fig.~\ref{fig:5}.
%
%
A comparison of Figs.~\ref{fig:1}(b) and \ref{fig:5}(a) shows that
Kondo screening of the impurity strongly suppresses the DOS of the
zero-energy eigenmode at the impurity position, $\rR$, and its
mirror site $-\rR$. The same result holds for eigenmodes away from
zero energy [see, e.g. Figs.~\ref{fig:1}(c) and \ref{fig:5}(b)]. In
contrast, the DOS of the Kondo resonance at $\omega=-0.019E_0$
possesses a peak around $\rR$, with a weaker image at the mirror
site, $-\rR$, while the resonance at $\omega=0.018E_0$ only exhibits
a peak in the DOS at the mirror site, but not at $\rR$. We find that
similarly strong signatures of Kondo screening in the DOS exist for
all locations of the magnetic impurity with $T<T_K({\rR})$.

Finally, results similar to the ones discussed above are also found
in other corral geometries and for different spatial structures of
the low-energy eigenmodes. Moreover, we expect that the spatially
dependent Kondo effect is robust against the inclusion of an
electron-electron interaction in the host system as long as the
relevant eigenmodes are located near the Fermi energy.

In summary, we studied the Kondo screening of a magnetic impurity
inside a non-magnetic quantum corral. We showed that the spatial
structure of the corral's low-energy eigenmode leads to spatial
variations in the Kondo temperature and the critical coupling. The
spectroscopic signature of the Kondo effect are multiple Kondo
resonances in the DOS with distinct spatial patterns. Our results
show that quantum corrals provide a new possibility to explore and
manipulate the Kondo effect.

The authors would like to thank R. Nyberg, A. Rosch, C. Slichter,
and M. Vojta for helpful discussions. D.K.M. acknowledges financial
support by the Alexander von Humboldt Foundation, the National
Science Foundation under Grant No. DMR-0513415 and the U.S.
Department of Energy under Award No. DE-FG02-05ER46225.

\vspace{-0.1cm}



\vspace{-0.1cm}

\begin{thebibliography}{99}
%
\bibitem{Kon64} J. Kondo, Prog. Theor. Phys. {\bf 32}, 37 (1964).

\bibitem{Wil75} K. G. Wilson, Rev. Mod. Phys. {\bf 47}, 774 (1975);
A.M Tsvelick and P.B. Wiegmann, Adv. Phys. {\bf 32}, 453 (1983); N.
Andrei {\it et al.}, Rev. Mod. Phys. {\bf 55}, 331 (1983).

\bibitem{Hew97} A. D. Hewson, {\it The Kondo Problem to Heavy Fermions}
 (Cambridge University Press, Cambridge, 1997).

 \bibitem{dots} D. Goldhaber-Gordon {\it et al.}, Nature (London) {\bf 391}, 156
(1998); S. M. Cronenwett, T. H. Oosterkamp, and L. P. Kouwenhoven,
Science {\bf 281}, 540 (1998).

\bibitem{tubes} M. R. Buitelaar {\it et al.}, Phys. Rev. Lett. {\bf 88}, 156801 (2002); J.
Nygard {\it et al.}, Nature (London) {\bf 408}, 342 (2000).

\bibitem{Boo05} C. H. Booth {\it et al.}, Phys. Rev. Lett. {\bf 95}, 267202
(2005).

\bibitem{Thi99} W.B. Thimm, J. Kroha, and J. von Delft, Phys. Rev. Lett. {\bf 82},
2143 (1999)

\bibitem{Sch02} For a review, see M. Pustilnik and L.I. Glazman, J. Phys.: Condens. Matter {\bf 16},  R513 (2004);
J. P. Schlottmann, Phys. Rev. B {\bf 65}, 024420 (2002); P.S.
Cornaglia and C.A. Balseiro, Phys. Rev. B {\bf 66}, 174404 (2002).

\bibitem{Cro95} M.F. Crommie {\it et al.}, Physica D {\bf 83}, 98 (1995).

\bibitem{Man00} H.C. Manoharan, C.P. Lutz, and D.M. Eigler, Nature (London) {\bf 403}, 512 (2000).

\bibitem{theory} G.A. Fiete {\it et al.}, Phys. Rev. Lett. {\bf 86}, 2392 (2001);
 A.A. Aligia, Phys. Rev. B {\bf 64}, 121102
(2001); K. Hallberg, A.A. Correa, and C.A. Balseiro, Phys. Rev.
Lett. {\bf 88}, 066802 (2002); D. Porras {\it et al.}, Phys. Rev. B
{\bf 63}, 155406 (2001); O. Agam and A. Schiller, Phys. Rev. Lett.
{\bf 86}, 484 (2001); Y. Shimada {\it et al.}, Surf. Sci {\bf 514},
89 (2002); M. Weissmann and H. Bonadeao, Physica E {\bf 10}, 544
(2001); M. Schmid and A.P. Kampf, Ann. Phys. {\bf 12}, 463 (2003);
for a general review see G.A. Fiete and E.J. Heller, Rev. Mod. Phys.
{\bf 75} 933 (2003), and references therein.

\bibitem{Read83}
 N. Read and D. Newns, J. Phys. C {\bf 16}, 3273 (1983).

\bibitem{Bic87} N. E. Bickers, Rev. Mod. Phys. {\bf 59}, 845 (1987).

\bibitem{Morr04}
 D. K. Morr and N. Stavropoulos, Phys. Rev. Lett. {\bf 92} 107006 (2004);
 {\it ibid.}, Phys. Rev. B {\bf 67} 020502(R) (2003).

\bibitem{com1} Note that the second term on the left hand side of
Eq.(\ref{eq:saddle1}) arises from $n_F(\epsilon)=1/2+T\sum_n
(i\omega_n - \epsilon)^{-1}$.

\bibitem{com2}  Note that $J=J_{cr}$ corresponds to the
solution $s^2,\epsilon_f=0$ of Eqs.(\ref{eq:saddle1}) and
(\ref{eq:saddle2}).

\bibitem{Ali05} A. A. Aligia and A. M. Lobos, J. Phys. Condens. Matter. {\bf 17}, S1095
(2005).


\end{thebibliography}
\end{document}